# On the effect of boundary vibration on Mucus Mobilization

Abiola D. Obembe[1], Mohammad Roostaie[2], Richard Boudreault[3], Yuri Leonenko[1,4,*]

## ABSTRACT

In this study, we investigate the mobilization of mucus in a cylindrical tube of constant cross-section subjected to a Small Amplitude Oscillatory Shear (SAOS) assuming the viscoelastic behavior of mucus is described by the Oldroyd-B constitutive equation. The Laplace transform method was adopted to derive expressions for the velocity profile, average velocity, instantaneous flowrate and mean flowrate (i.e., mobilization) of the mucus within the tube. Additionally, a 2D finite element model (FEM) was developed using COMSOL Multiphysics software to verify the accuracy of the derived analytical solution considering both a Newtonian and an Oldroyd-B fluid. Furthermore, sensitivity studies were performed to evaluate the influence of the vibration frequency, vibration amplitude, mean relaxation time, and zero-shear viscosity on mucus mobilization. Results prove that the analytical and numerical results are within acceptable error tolerance and thus in excellent agreement. Besides, the parametric studies indicate a 48%, 57%, and 343% improvement in mucus mobilization when the mean relaxation time, vibration amplitude, and vibration frequency, respectively were each increased by a factor of 6 from the assumed nominal values. Conversely, the mucus mobilization decreased by 25% when the zero-shear viscosity was increased by a factor of 6 from the assumed reference value. In general, this study confirms that mucus mobilization in a tube can be improved by increasing the magnitude of vibration amplitude and vibration frequency. Similarly, the larger the magnitude of mucus mean relaxation time the better the mucus mobilization when the tube is subjected to boundary vibrations. Finally, mucus mobilization decreases as the magnitude of mucus viscosity increases.

[1] Department of Earth and Environmental Science, University of Waterloo, Waterloo ON, Canada, N2L 3G1
[2] Department of Mechanical Engineering, University of Toronto, Toronto, ON, Canada, M5S 1A1
[3] Techaero, Montreal, QC, H8T 3H7
[4] Department of Geography and Environmental Management, University of Waterloo, Canada, N2L 3G1
*Corresponding author (leonenko@uwaterloo.ca)

## 1. INTRODUCTION

The effect of mechanical oscillations on the flow of viscoelastic and non-Newtonian fluids has garnered the attention of a plethora of researchers due to its numerous application in different industrial processes including pipe flow, secondary oil recovery, filtration, and fluid pumping [1–7]. Adequate understanding of the complex flow dynamics in these industrial processes is key to preventing operational challenges and optimizing flow processes (i.e., reducing total power requirements, and the detrimental effects induced by sudden start-up or shut down during pipe flow). It is argued that varying the pump or valve pulse frequencies in pipes helps alleviate problems related to sudden start-up or shut down of pipe flows [8–11]. Other important applications of oscillatory fluid flow comprise acoustics propagation in biological organs (i.e., in the respiratory and circulatory systems of living beings).

Understanding the biological-fluid dynamics, including heartbeat, stenosis, and mucus movement, is beneficial in the diagnosis and treatment of diseases, including cardiovascular and respiratory diseases, and the design of new mechanical systems associated with the biological systems in our bodies [12]. A potentially detrimental example of cardiovascular disease is the sudden blockage resulting from blood clots increasing blood pressure. The resulting increase in pressure typically leads to intracerebral hemorrhage due to the high pressure and shear stress on the vessels' walls upstream of the blockage site and the oxygen deprivation on the downstream side [13]. Therefore, an in-depth understanding of the atherosclerotic-plaques formation in carotid arteries and the associated shear stress in blood vessels is crucial for medical research, diagnosis, and treatment planning [14–16]. Similarly, Mucus is acknowledged to exhibit non-Newtonian viscoelastic properties, and numerous studies have been devoted to the application of acoustic waves and mechanical oscillations to address respiratory airway clearance. Specifically, the mucus adherence and accumulation in the alveoli of COVID-19 patients exacerbate the already declining lung function by airflow reduction, airway obstruction, and respiratory failure, initiating COVID-19-induced mortality [17–19]. Some of the devices created to achieve improved airway clearance have also been successfully deployed to treat cystic fibrosis, pulmonary, neurological, and neuromuscular diseases [20–22].

In general, non-Newtonian fluids may exhibit either finite yield stress, or shear-thinning, or thixotropy, or a combination of all these characteristics [23–25]. It has been presented in numerous studies that the shear modulus of some of these fluids is frequency-dependent and increases to a plateau as the oscillation frequency increases [26–28]. Stress application rate and its continuity, triggered by acoustic waves, alter the non-linear stress-strain curve of non-Newtonian fluids and their elasticity, resulting in an improved displacement of these fluids[29]. The mathematical literature in this field can be categorized into either analytical or numerical approaches. Analytical methods provide for rapid prototypes and better mechanistic/physical insight into the phenomena. Alternatively, numerical methods provide for a more comprehensive and robust investigation of physical problems since they require fewer assumptions in their development.

In 1963, Ting [30] derived the first exact solution of the non-Newtonian second-grade fluids flow in a cylinder. Subsequently, Srivastava [31], and Waters and King [32] provided the exact solutions, respectively for Maxwell and Oldroyd-B fluids. The Maxwell constitutive relationship was presented as a simple mechanical model for describing viscoelastic materials that comprise a linear spring and a linearly viscous dashpot arranged in series [33,34]. Similarly, the Oldroyd-B model is popular in modeling non-linear fluids and is a three-dimensional generalization of the Maxwell model. This model takes into account the fluid incompressibility and its response when satisfying appropriate invariance requirements [35]. Rajagopal [36] presented the exact solutions for both an incompressible and homogeneous second-grade fluid and studied the effects of torsional and longitudinal oscillations in an infinite rod. Rajagopal and Bhatnagar [37] extended their original solutions for an Oldroyd-B fluid. Penton [38] studied the flow of a viscous fluid initiated by an oscillating plate and provided the first closed-form transient solution for this problem. Erdogan [39] using the Laplace transformation technique derived two starting solutions for the flow of a linearly viscous fluid based on the cosine and sine oscillations of a flat plate. Their solution was further extended to non-Newtonian fluids using the second-grade fluid, Maxwell, and Oldroyd-B models in the following literature just to name a few [37,40–51].

Theoretical predictions show that perturbation methods on viscometric flows (or nearly viscometric flows) can increase the flow rate as a function of frequency and amplitude of oscillations [4,5,52–59]. Mena et al. [4] using numerical and experimental procedures showed that a longitudinal vibration of the pipe wall strongly affects the flow in non-Newtonian fluids. The viscoelastic properties, shear-thinning behavior, and amplitude of the superimposed oscillations are the main parameters of this phenomenon with elasticity as a secondary parameter. Herrera-Velarde et al. [57] investigated the temperature change of a non-Newtonian fluid due to the oscillation of the pipe wall. Their results illustrate peaks in flow rate enhancement at specific wall shear stresses. Fetecau et al. [60] adopted the Fourier transform method to analytically investigate the effect of sidewalls on the flow of a viscous fluid over an oscillating infinite plate. Chen et al. [61] employed experimental and analytical methods to investigate the effect on the flow behavior of a cylindrical rod vibrating in a cylindrical shell filled with viscous fluid. Duarte et al. [62] investigated the flow of a viscoelastic fluid both in pulsating and start-up tests using upper-convected Maxwell and Oldroyd-B fluid models. Kazakia and Rivlin [3,63] in two consecutive analytical studies investigated the effects of longitudinal, transverse, and rotational vibrations and their combination on non-Newtonian fluid flow in a pipe. They showed that when the fluid viscosity decreases due to the increase in the shear rate, the discharge rate of non-Newtonian fluid may increase with the vibration frequency and amplitude. The fluid flow in their works is assumed to be laminar. Phan-Thien [2] also provided an analytical method to investigate the effect of vibration on fluid flow. On the other hand, recent studies suggest employing specific fluid models, such as upper-convected Maxwell and Oldroyd-B, for non-Newtonian fluids. Hullender[7] studied analytically the pre-transient and turbulent flow of an Oldroyd-B fluid in a circular pipe.

Surprisingly, the effects of vibration on the mucus movement, whether it is in conjunction with airflow or not, have not still been closely examined. Chang et al. [64] in their experimental and theoretical studies about mucus-air movement in airways demonstrated the significant effect of shear stress at the air-mucus interface on mucus velocity. Button et al. [65] reported the effectiveness of cyclic compressive stress, chest percussion, and mechanical therapeutic devices in lung clearance of cystic fibrosis patients. In 1982, Radford et al. [66] reported altered mucus flow rates by applying percussion energy to dogs' and humans' chest walls with the optimal frequency range at 25-35 Hz. In 1983, King et al. [67,68] studied the mucus clearance in the trachea of nine anesthetized dogs using high-frequency chest wall compression. In their work, the tracheal mucus clearance rate increased with the most pronounced clearance in the range of 11-15 Hz oscillation, peaking at 13 Hz. Then, Gross et al. [69] also reported improvements in peripheral mucociliary clearance by applying high-frequency chest wall oscillation at a 13 Hz and a similar spontaneously breathing population and measurement technique in anesthetized dogs. Rubin et al. [70] experimentally found that a low-energy chest wall oscillator (13 Hz) improved mucus clearance in central airways compared to a commercial chest percussor (40 Hz) in five anesthetized dogs. Ragavan et al. [71] identified the significant interactive impact of tracheal angles, cough velocities, oscillations, and simulant types on airway clearance during cough. The mucus simulant with high elasticity and cohesion displaced significantly longer distances at all cough velocities, at all angles of tracheal inclination with or without airflow oscillations compared with those of thinner mucus simulant. Furthermore, superimposed flow oscillations of 25-68 Hz had a significant influence on increasing the displacement of different types of mucus simulant. The current mathematical investigations and studies of airway clearance using high-frequency oscillation on chest walls require further development to better understand the mechanisms behind this phenomenon. This understanding is crucial and greatly beneficial to the COVID-19 patients, particularly because these treatment strategies are non-surgical and non-pharmaceutical.

In the present work, we aim to gain an in-depth understanding of the linear viscoelastic response of mucous in a cylindrical pipe driven by a constant pressure gradient and superposed by mechanical oscillations/acoustics in the longitudinal direction. The pipe is considered to have a constant cross-section and the walls are perturbed by a mechanical oscillation(s) at a specified frequency and amplitude. A 1D semi-analytical model is derived for this purpose, adopting the Oldroyd-B constitutive equation model for mucous and employing the Laplace transform method to derive semi-analytical expressions for the velocity field, average velocity, and volumetric flow rate of the mucus. Laplace transformation plays a fundamental role in modeling and analyzing numerous engineering design problems because it renders the problem of solving a linear constant-coefficient differential equation to solving an algebraic equation that is easier to handle[72][73]. Furthermore, a 2D finite element model (FEM) is simultaneously developed using the viscoelastic module of the COMSOL 5.6 Multiphysics software to verify the accuracy of the semi-analytical solution. Subsequently, the influence of the relaxation time, polymer viscosity, vibration frequency, and amplitude, on mucus mobilization are both qualitatively and quantitatively examined by parametric studies. The results of this work will provide initial insights into the order of magnitude of mucus mobilization predicted by the Oldroyd-B linear viscoelastic constitutive equation as well as illustrate the transient response of mucous in vibrated media. To the best of our knowledge, none of the previous works has analytically studied the transport of mucus when mechanical acoustics are superposed on the walls. Therefore, we argue that this work can pave the way to designing devices that may help patients suffering from an accumulation of mucus in the lungs due to COVID-19.

The agenda of this paper is enumerated as follows: Section 2 presents an overview of the flow problem and step-by-step development of the mathematical model in the radial-cylindrical coordinate system. The list of underlying assumptions required to arrive at the analytical solution and the expressions for the axial velocity, average velocity, and volumetric flow rate are presented in Section 3. Besides, the finite element model (FEM) and mathematical equations numerically solved in COMSOL Multiphysics software are detailed in the remainder of Section 3. Section 4 comprises two main sections; In Section 4.1, the semi-analytical solution is verified against the FEM model (Section 4.1) and in Section 4.2, the results of the parametric study conducted are discussed in detail. The paper ends with some conclusions and remarks in Section 5.

## 2. THEORETICAL FORMULATION

### 2.1. Derivation of Mathematical Model

In the following, we present the theoretical description of the transient flow problem of mucous within a vibrated tube where the background flow is driven by a constant pressure gradient. Typically, such vibrations at the pipe wall, defined herein by $u_W(t)$, induce mechanical oscillation in the bulk fluid and micro-structures encountered in such rheological complex fluids. In reality, wall vibration in human flow passages may be triggered through a mechanical solicitation (e.g., clapping of the chest walls)[74] or mechano-acoustic treatment devices (i.e. acoustic percussion)[29], to mention but a few. Figure (**Fig.**) 1 shows the schematic of the pipe with length $L$ and radius $R$ where the pressures at the ends of the pipe are denoted by $P_A$ and $P_B$ are indicated.

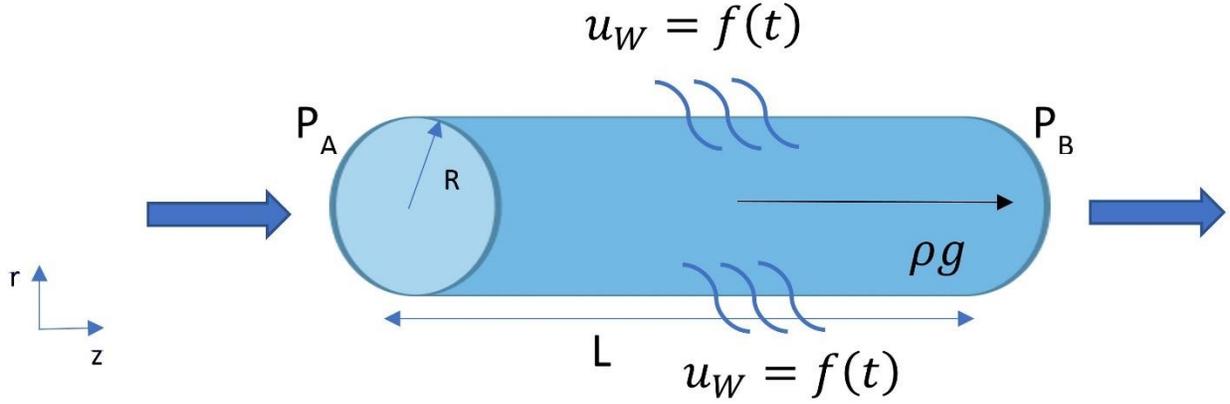

**Fig. 1** Schematic of the flow geometry

Therefore, the fluid transients in the pipe stem from the net pressure drop ($\Delta p$) in the pipe stemming from the pressure at the ends of the pipe and pressure due to gravity. The governing equations for non-Newtonian flow problems comprise the momentum equation (Navier Stoke), continuity equation, and an appropriate constitutive relationship between stress and shear rate.

The Navier-Stokes and continuity equations in the vectorial form are expressed as follows:

$$\rho \left( \frac{\partial u}{\partial t} + u.\nabla u \right) = -\nabla p + \nabla.\sigma + \rho g \tag{1}$$

where $p$ represents pressure and $\rho$ is the mucous density and $g = 9.81$ m/s$^2$ is the gravitational acceleration.

$$\frac{\partial \rho}{\partial t} + \nabla.(\rho u) = 0 \tag{2}$$

Therefore, considering the cylindrical geometry of the pipe, the governing equations describing the laminar motion of viscoelastic gel under unsteady-state condition comprise [7]:

$$\rho \frac{\partial u_z}{\partial t} = \psi + \frac{1}{r} \frac{\partial (r\sigma)}{\partial r} \tag{3}$$

$$\frac{\partial u_z}{\partial z} = 0 \tag{4}$$

where $u_z$ is the axial velocity, $\psi = \rho g + \frac{P_A - P_B}{L}$ or $\psi = \rho g$ for flow driven purely by gravity is the net constant pressure gradient, $\sigma$, and $\rho$ define the axial velocity, net pressure gradient, stress tensor, and fluid density respectively.

Besides, $u_z$ is a function of $r$, $\sigma$ is a function of $r$ and $z$; however, $p$ is assumed to be uniform at any specific $z$. Additionally, to aid the mobilization of mucous within the tube, longitudinal oscillations are introduced at the pipe wall. The velocity of the wall is given by the following relation:

$$u_W(t) = u_z(r,t)_{r=R} \tag{5}$$

where $u_W(t)$ is a general function of time that allows for easy Laplace transformation.

Specifically, we consider two different expressions for the wall velocity in this research:

$$u_W(t) = b_0 t \tag{6}$$

$$u_W(t) = A\omega \cos(\omega t + b_1) \tag{7}$$

Here, $b_0$ is the acceleration coefficient, $b_1$ is the phase angle, $A$ is the displacement amplitude and $\omega = 2\pi f$ is the angular frequency of the vibration, and $f$ is the frequency of vibration. The choice of including the constant $b_1$ (i.e., phase shift) in Eq. (7) is to cater for periodic functions or acoustic waves that do not necessarily start at the sinusoidal axis or a maximum or a minimum. Specifically, this allows periodic functions with the same amplitude and angular frequency to exhibit different starting points.

Finally, we assume the stress strain-rate relationship for the fluid is described by the Oldroyd-B tensor model[75] expressed as follows:

$$\left[\sigma + \lambda_1 \frac{\partial \sigma}{\partial t}\right] = \mu_0 \left[\frac{\partial u_z}{\partial r} + \lambda_2 \frac{\partial^2 u_z}{\partial r \partial t}\right] \tag{8}$$

Here $\mu_0$ is the zero-shear viscosity which quantifies the viscous contribution, $\lambda_1$ is the relaxation time and $\lambda_2 = \frac{\mu_s}{\mu_0} \lambda_1$ is the retardation time. Besides, Eq. (8) reduces to the Maxwell constitutive equation when $\lambda_2 = 0$, and $G = \frac{\mu_0}{\lambda_1}$ is the elastic modulus which quantifies the elastic contribution.

## 3. MATHEMATICAL MODEL AND ITS SOLUTIONS

### 3.1. Analytical Solution

In this section, the semi-analytical solution for the mathematical model developed in Section 3.1 is detailed in the Laplace domain. Interested readers may refer to Appendix A for complete derivation.

The following assumptions are considered to arrive at an analytical solution for laminar flow of mucus in a vertical smooth tube with a uniform cross-sectional area:

a) Flow is axisymmetric, and the fluid is assumed to be incompressible.

b) Isothermal fluid flow

c) The radial and azimuthal components of the fluid velocity are zero (i.e., purely axial flow) and thus neglect the pressure gradient and velocity components normal to the wall of the tube. Therefore, the contribution of the nonlinear term in Eq. (1) is negligible.

d) The viscoelastic behavior of mucus is described by the Oldroyd-B constitutive equation.

e) The mucus viscosity is assumed to be constant i.e., shear-thinning viscosity is neglected.

### 3.1.1. Velocity profile

The transient flow field (i.e., velocity profile) within the tube in the Laplace domain given by Eq. (A-25) in Appendix A is:

$$\widehat{U}_z = a_1 \mathrm{I}_0(cr) + \frac{\widehat{\psi}}{\rho s} \tag{9}$$

Where the coefficient $a_1$ is defined in Eq. (A-26) as:

$$a_1 = \frac{1}{\mathrm{I}_0(cR)}\left(\widehat{U}_W - \frac{\widehat{\psi}}{\rho s}\right) \tag{10}$$

Substitution of Eq. (10) in Eq. (9) leads to:

$$\widehat{U}_z = \frac{\mathrm{I}_0(cr)}{\mathrm{I}_0(cR)}\widehat{U}_W + \frac{\widehat{\psi}}{\rho s}\left[1 - \frac{\mathrm{I}_0(cr)}{\mathrm{I}_0(cR)}\right] \tag{11}$$

where $\widehat{U}_W$ is defined as the Laplace transform of $u_W$.

### 3.1.2. Average velocity

Assuming the flow passage is a circular tube with radius, $R$, the average velocity is calculated by the expression:

$$\widehat{U}_{av} = \frac{1}{\pi R^2}\int_0^R 2\pi r \widehat{U}_z \, dr \tag{12}$$

Substituting Eq. (9) in Eq. (12) results in

$$\widehat{U}_{av} = \frac{2}{R^2}\left[a_1 \int_0^R r\mathrm{I}_0(cr)\, dr + \frac{\bar{\psi}}{\rho s}\int_0^R r\, \mathrm{dr}\ \right] \tag{13}$$

Adopting integration by parts, Eq. (13) reduces to

$$\widehat{U}_{av} = \frac{2}{R^2}\left[\frac{a_1 R}{c}\mathrm{I}_1(cR) + \left(\frac{\bar{\psi}}{\rho s}\right)\frac{R^2}{2}\right] \tag{14}$$

For convenience, Eq.(14) can be expressed as

$$\widehat{U}_{av} = \left[\frac{2a_1}{cR}\mathrm{I}_1(cR) + \frac{\bar{\psi}}{\rho s}\right] \tag{15}$$

### 3.1.3. Instantaneous and mean flowrate

We define the instantaneous flowrate in the tube by the expression:

$$\widehat{Q} = 2\pi \int_0^R r\widehat{U}_z\, dr \tag{16}$$

Similarly, the substitution of Eq. (9) in Eq. (16) yields

$$\widehat{Q} = 2\pi \left[a_1 \int_0^R r\mathrm{I}_0(cr)\, dr + \frac{\bar{\psi}}{\rho s}\int_0^R r\, \mathrm{dr}\ \right] \tag{17}$$

Performing integration by parts, the instantaneous flowrate is governed by the expression:

$$\widehat{Q} = \pi \left[\frac{2a_1 R}{c}\mathrm{I}_1(cR) + \frac{\bar{\psi}\, R^2}{\rho s}\right] \tag{18}$$

Similarly, we define the mean flowrate as the average of the instantaneous flowrate over a period of oscillation in the real-time domain by the expression[76]:

$$Q_m = \frac{1}{T}\int_0^T Q(t)dt \tag{19}$$

Where $T$ is the period of oscillation.

Applying the Laplace transform operator ($\mathcal{L}$) on Eq. (19) leads to:

$$\widehat{Q}_m = \frac{1}{T}\left(\frac{\widehat{Q}}{s}\right) \tag{20}$$

Where $\widehat{Q}$ is the instantaneous volumetric flowrate in Laplace space derived from Eq. (18).

### 3.1.4. Numerical inversion

The expressions for velocity, average velocity, and instantaneous flowrate have been converted from the Laplace-domain to the real-time domain using the Sthefest algorithm for Laplace transform inversion is computed by the expression[77]:

$$f(t) \sim [s \cdot \sum_{n=1}^{N} K_n \cdot F(ns)]_{s=\ln(2/t)} \tag{21}$$

And the weighting coefficients $K_n$ are defined by

$$K_n = (-1)^{n+\frac{N}{2}} \cdot \sum_{k=(n+1)/2}^{minimum(n,N/2)} \frac{k^{\frac{N}{2}}(2k)!}{(N/2-k)!k!(k-1)!(n-k)!(2k-n)!} \tag{22}$$

All computations in this research were implemented using MATLAB scientific programming language.

### 3.2. Finite Element Solution

A 2D numerical model was simultaneously implemented using the viscoelastic flow module in the COMSOL Multiphysics software (version 5.6) to validate the herein derived semi-analytical solution. The tube is modeling using the 2D axisymmetric geometry (see **Fig. 2**) and the fluid is treated as a viscoelastic gel.

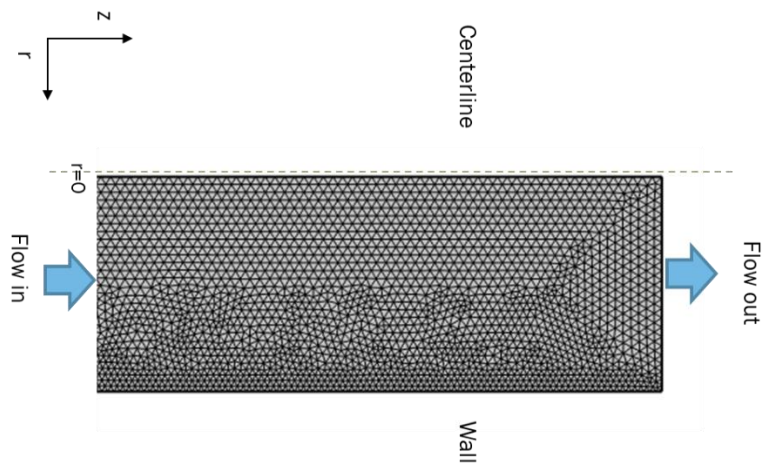

**Fig. 2** Meshed 2D axisymmetric geometry of the vibrated tube

Additionally, the tube wall is specified with the appropriate condition, and the pressure at the inlet and outlet of the tube is specified as $P_A$ and $P_B$, respectively. The program solves the following coupled equations [78]:

$$\rho \frac{\partial u}{\partial t} + \rho(u.\nabla)u = \nabla.\left[-pI + K + \sigma\right] \tag{23}$$

$$\rho \nabla.u = 0 \tag{24}$$

$$K = \mu_s[\nabla u + (\nabla u)^T] \tag{25}$$

$$\sigma = \sum_m \sigma_m \tag{26}$$

$$\lambda_m \vec{\sigma}_m + \sigma_m = 2\mu_{pm}S \tag{27}$$

$$\vec{\sigma}_m = \frac{\partial \sigma_m}{\partial t} + (u.\nabla)\sigma_m - e \cdot \sigma_m - \sigma_m \cdot e^T \tag{28}$$

$$S = \frac{1}{2}(e + e^T), \; e_{ij} = \frac{\partial u_i}{\partial x_j} \tag{29}$$

Where $m$ is the number of modes/branches of the Oldroyd-B model.

It is important to highlight that the FEM model easily accommodates the shear-thinning viscosity of the gel (i.e., power-law model, Carreau-Yasuda, etc.)

### 3.3.    Summary

In this paper, two different solutions for the flow of a viscoelastic fluid within a vibrated tube are presented. Both solutions neglect the shear-thinning property of the viscosity; however, more nonlinear viscoelastic models such as the Giesekus model can easily be implemented using a FEM model to cater to the shear-thinning property of viscosity. **Table 1** provides an overview of the key attributes of both solutions presented in this research.

**Table 1** Characteristics of analytical and numerical solutions

| Solution | Shear-Thinning | Viscoelasticity |
|---|---|---|
| Analytical | No | Yes |
| Oldroyd-B FEM | No | Yes |

## 4. RESULT AND DISCUSSION

The results and discussion section comprises of the following: In Section 4.1, the derived semi-analytical solution is validated against the FEM solution obtained from the COMSOL Multiphysics software for steady-state and transient flow synthetic problems. Additionally, the remainder section (Section 4.2) which consists of four sub-sections presents the results of the parametric analysis conducted to illustrate the effects of the vibration frequency, vibration amplitude, mean relaxation time, and zero-shear viscosity on the flow and mucus mobilization within the vibrated tube.

### 4.1.    Validation of Solutions

Herein, the accuracy of the derived analytical solution for flow field in the vibrated cylindrical tube was verified against the numerical model from COMSOL considering both the static and dynamic flow configurations. First, a stationary analysis is computed with no movement of the pipe wall considering both a vicious and viscoelastic fluid. Furthermore, a time-dependent analysis for a viscoelastic fluid is solved where the tube wall velocity is prescribed according to Eq. (6). Besides, to establish the accuracy of the analytical solution we neglect the shear-thinning of viscosity in the FEM model. **Table 2** enumerates the synthetic base-case input data utilized for validating the analytical and numerical solutions.

**Table 2** Synthetic material properties used in verification of solutions

| | |
|---|---|
| $r_o = 2 \times 10^{-3}$m | $t_{end} = 0.005$ s |
| $\mu_p = 10$ Pa·s | $\mu_s = 1 \times 10^{-3}$ Pa·s |
| $\psi = 100$ Pa·m$^{-1}$ | $b_1 = 0$ |
| $\rho = 1000$ kg·m$^{-3}$ | $\lambda_1 = 0.1$ s |
| $\mu_0 = \mu_p + \mu_s = 10.0010$ Pa·s | $\rho = 1000$ kg·m$^{-3}$ |
| $\lambda_2 = \frac{\mu_s}{\mu_0} \lambda_1 = 10 \times 10^{-6}$s | $L = 50 \times 10^{-3}$ m |
| $b_0 = 1 \times 10^{-6}$ m/s$^2$ | |

#### 4.1.1.  Steady-State Solution

The steady-state case was first solved considering a viscous fluid and a viscoelastic fluid for the limiting case of no wall movement (i.e., $u_W = 0$). It is important to note that setting $\lambda_1 = \lambda_2$, the Oldroyd-B constitutive equation reduces to a Newtonian fluid with a viscosity equal to the solvent viscosity $\mu_s$. Noting that the zero-shear viscosity is the addition of the polymer viscosity $\mu_p$ and $\mu_s$ implies that $\mu_p = 0$ for a Newtonian fluid. Furthermore, to arrive at the steady-state solution, we set $t \to \infty$ for the analytical solution. The velocity profiles generated from both solutions considering a viscous fluid and a viscoelastic gel are shown in **Fig. 3**.

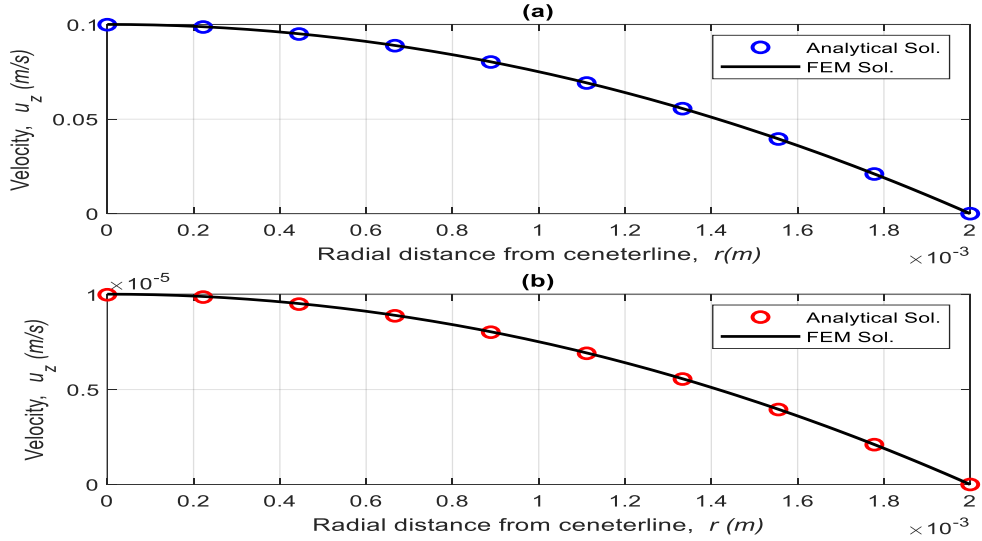

**Fig. 3** Axial velocity profile solution in the stationary regime for (a) Newtonian fluid and (b) Viscoelastic fluid

Each data point in the plot corresponds to the velocity at a given position along the pipe radius at steady-state conditions. As expected, **Fig. 3**(b) indicates that the velocity profile for the viscoelastic elastic fluid is $\sim 10^4$ order of magnitude smaller than that of the Newtonian fluid in **Fig. 3**(a). Besides, an inspection of **Fig. 3** reveals that there is reasonable overlap between both curves for both fluids providing a first-step verification of the derived analytical solution and numerical model. Additionally, the average velocity and volumetric flow rate predicted from both solutions are in close agreement with a relative error of less than 1% as shown in Table 3.

**Table 3** Quantitative comparison of solution for the steady-state problem

| Solution | Analytical | | FEM | |
|---|---|---|---|---|
| Fluid | Average velocity (m/s) | Volumetric flow rate (m³/s) | Average velocity (m/s) | Volumetric flow rate (m³/s) |
| Newtonian | 0.05 | $6.28 \times 10^{-7}$ | 0.0501 | $6.29 \times 10^{-7}$ |
| Viscoelastic | $5 \times 10^{-6}$ | $6.28 \times 10^{-11}$ | $5.02 \times 10^{-6}$ | $6.30 \times 10^{-11}$ |

*4.1.2.  Unsteady State Solution with prescribed wall velocity*

The accuracy of the derived analytical solution was further verified against the numerical solution by performing a time-dependent analysis for an Oldroyd-B fluid where the tube wall velocity is defined as a linear function of time i.e., Eq. (6). Applying the Laplace transform operator on the Eq. (6) leads to:

$$\hat{U}_w = \frac{b_0}{s^2} \tag{30}$$

Adopting a linear function of time for the wall velocity, the flow behavior (i.e., velocity and flow rate) of the viscoelastic fluid predicted from both solutions are displayed in **Fig. 4.**

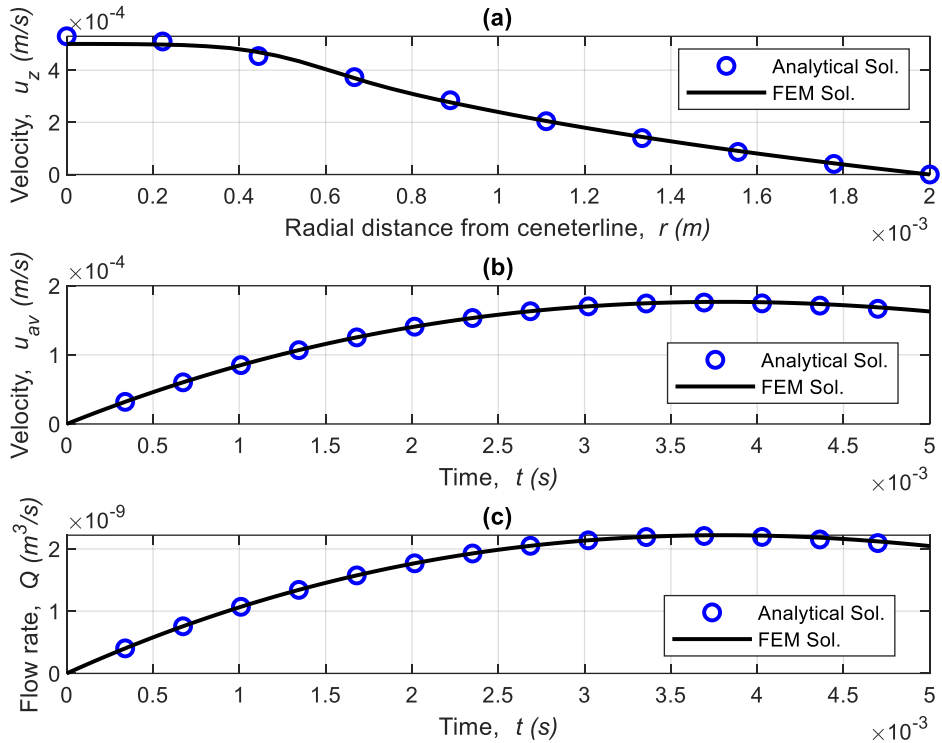

**Fig. 4** Model validation for a dynamic wall (a) axial velocity profile at $t = 0.005$ s, (b) average velocity history, and (c) volumetric flow rate history

The curves generated in **Fig. 4** portray the velocity profile at an elapsed time of 0.005 s, and the average velocity and volumetric flow rate, respectively. The excellent overlap of the curves in the plot establishes that there is a close agreement between the derived analytical solutions and the FEM solution.

### 4.2. Sensitivity Studies

Herein, we investigate the transient flow behavior of mucus through a vibrated tube where the flow is driven by a combination of pressure and gravity. In addition, the axial velocity at the wall of the tube is an oscillating function of time as defined in Eq. (7). The corresponding sinusoidal wall velocity in the Laplace domain is given by the expression:

$$\hat{U}_w = A\omega \left[\frac{s\cos(b_1) - \omega\sin(b_1)}{s^2 + \omega^2}\right] \tag{31}$$

Additionally, to capture the viscoelastic behavior of human sputum, rheometric data representative of the healthy range of human lung mucus extracted from the literature was utilized as input data[79]. **Table 4** enumerates the data and corresponding fit using five Upper Convected Maxwell modes.

**Table 4** Parameters for five-mode Giesekus model for a healthy range of human lung mucus[79]

| Mode | Relaxation time ($\lambda_1$), [s] | Elastic modulus ($G$), [Pa] | Giesekus parameter, α |
|------|------------------------------------|------------------------------|------------------------|
| 1    | 0.0089                             | 3.3472                       | 0.2                    |
| 2    | 0.0821                             | 0.7551                       | 0.3                    |
| 3    | 0.4660                             | 0.5350                       | 0.5                    |
| 4    | 3.1290                             | 0.4543                       | 0.5                    |
| 5    | 49.733                             | 0.8486                       | 0.5                    |

Accordingly, from the data listed in **Table 4**, the zero-shear viscosity and mean relaxation time can be estimated from the following expressions[80]:

$$\mu_0 = \mu_s + \sum_{i=1}^{5} G_i \lambda_{1_i} \tag{32}$$

$$\bar{\lambda}_1 = \frac{\sum_{i=1}^{5} G_i \left(\lambda_{1_i}\right)^2}{\sum_{i=1}^{5} G_i \lambda_{1_i}} \tag{33}$$

The contribution of the Giesekus parameter would be neglected in this study since the Oldroyd-B constitutive equation assumes the mucus viscosity is constant (i.e., shear-rate-independent). Besides, only the smallest relaxation time would be utilized to characterize the mucus. Additional data presented in **Table 5** would be utilized to quantitatively establish the effects of vibration frequency, vibration amplitude, relaxation time, and zero-shear viscosity on flow quantities (i.e., velocity and instantaneous flow rate, and mean flow rate).

**Table 5** Base-case data for parametric study

| | |
|---|---|
| $r_o = 2 \times 10^{-3}$ m | $P_A = 2$ Pa |
| $\mu_s = 1 \times 10^{-3}$ Pa·s | $\mu_p = 44$ Pa·s |
| $\mu_0 = \mu_s + \mu_p = 44.001$ Pa·s | $L = 20 \times 10^{-3}$ m |
| $b_1 = -\frac{\pi}{18}$ | $\rho = 1000$ kg·m$^{-3}$ |
| $f = 15$ Hz | $\bar{\lambda}_1 = 0.089$ s |
| $\bar{\lambda}_2 = \frac{\mu_s}{\mu_0}\bar{\lambda}_1 = 8.89 \times 10^{-7}$ s | $A = 1.6 \times 10^{-3}$ m |

To ensure a laminar flow regime, the parameters were chosen such that the vibrational Reynolds number ($Re_v = \frac{A\omega D}{v_e}$) was always less than 30 [81]. Finally, the transient vibrational simulation was conducted over 4 periods of the cosine wave (i.e., from $\omega t_0 = 0$ to $\omega t_{end} = 8\pi$).

### 4.2.1. Sensitivity on the vibration frequency

The magnitude of frequency denotes the number of vibrations (i.e., pressure peaks) perceived per second. The effect of frequency on the vibrational flow of mucus was investigated by independently varying the frequency between 15 Hz to 90 Hz while keeping other inputs at their base level enumerated in **Table 5**. As shown in **Fig. 5**, increasing the magnitude of vibration frequency leads to axial velocity curves shifted upwards (i.e., larger axial velocity values).

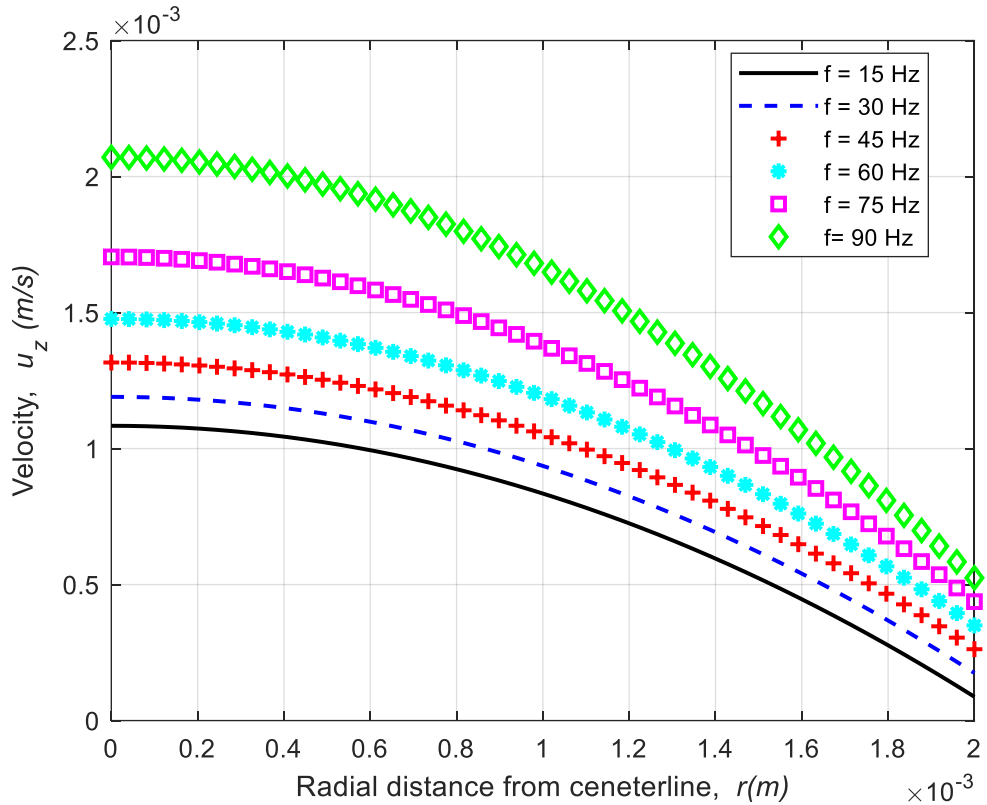

**Fig. 5** Effect of vibration frequency on the axial velocity profile

Thus, **Fig. 5** suggests that mucus transport within the tube improves as the magnitude of vibration frequency increases. To further establish this hypothesis, the average velocity, and instantaneous flow rate as a function of vibration frequency after four periods of the cosine wave are exhibited in **Fig. 6**.

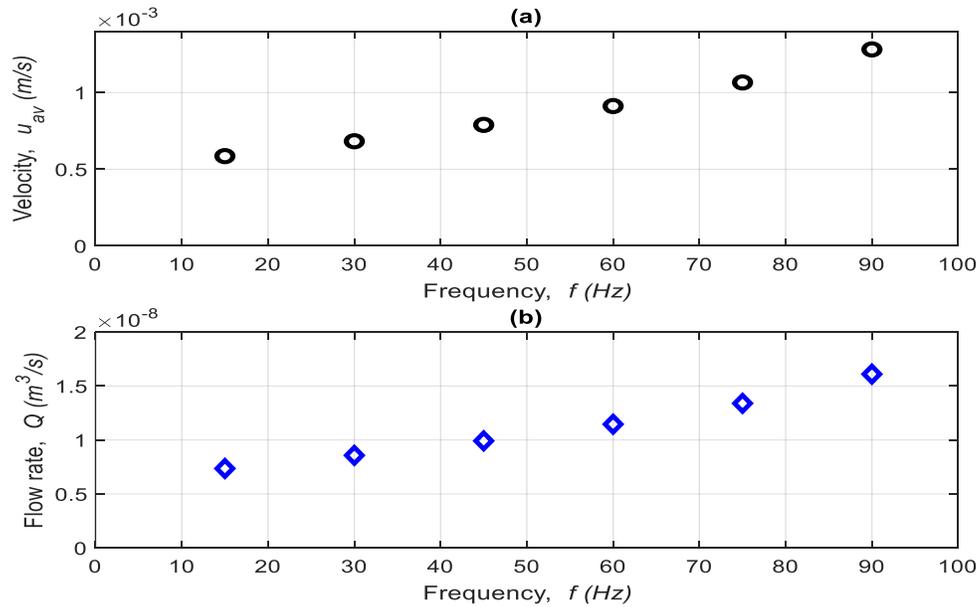

**Fig. 6** Effect of vibration frequency on (a) average velocity and (b) Instantaneous flowrate

Evidently, **Fig. 6** indicates that that the vibration frequency has a positive relationship with both the average velocity and instantaneous flow rate, respectively (i.e., the average velocity and instantaneous flow rate increase with vibration frequency). For instance, the instantaneous flow rate is $7.36 \times 10^{-9}$ m³/s, and $1.61 \times 10^{-8}$ m³/s, when $f = 15$ Hz, and 90 Hz, respectively. Furthermore, the effect of vibration frequency on the mean flow rate (i.e., mucus mobilization) was quantified and summarized in **Table 6.**

**Table 6** Mucus mobilization for different vibration frequencies

| Frequency (Hz) | Mean flowrate, $Q_m$ (m³/s) |
|----------------|------------------------------|
| 15 | $1.98 \times 10^{-8}$ |
| 30 | $3.33 \times 10^{-8}$ |
| 45 | $4.69 \times 10^{-8}$ |
| 60 | $6.05 \times 10^{-8}$ |

| 75 | $7.41 \times 10^{-8}$ |
|----|----------------------|
| 90 | $8.77 \times 10^{-8}$ |

Of interest, **Table 6** confirms that for a $A = 1.6 \times 10^{-3}$ m, when the vibration frequency increased by a factor of 2 (i.e., from 15 Hz to 30 Hz) and a factor of 6 (i.e., from 15 Hz to 90 Hz), the mucus mobilization improved by 68% and 343%, respectively. Consequently, we argue that a linear viscoelastic constitutive model can predict a 68% improvement in mucous mobilization by only increasing the vibration frequency by a factor of 2 i.e., from 15 Hz to 30 Hz.

### 4.2.2. Sensitivity on the vibration amplitude

The influence of the amplitude of vibration on the flow behavior including mucus mobilization was quantified by varying the amplitude values up to $1.6 \times 10^{-3}$ m while keeping other parameters fixed at their base values listed in **Table 5**. As shown in **Fig. 7**, the axial velocity curves are shifted upward as the vibration amplitude increases.

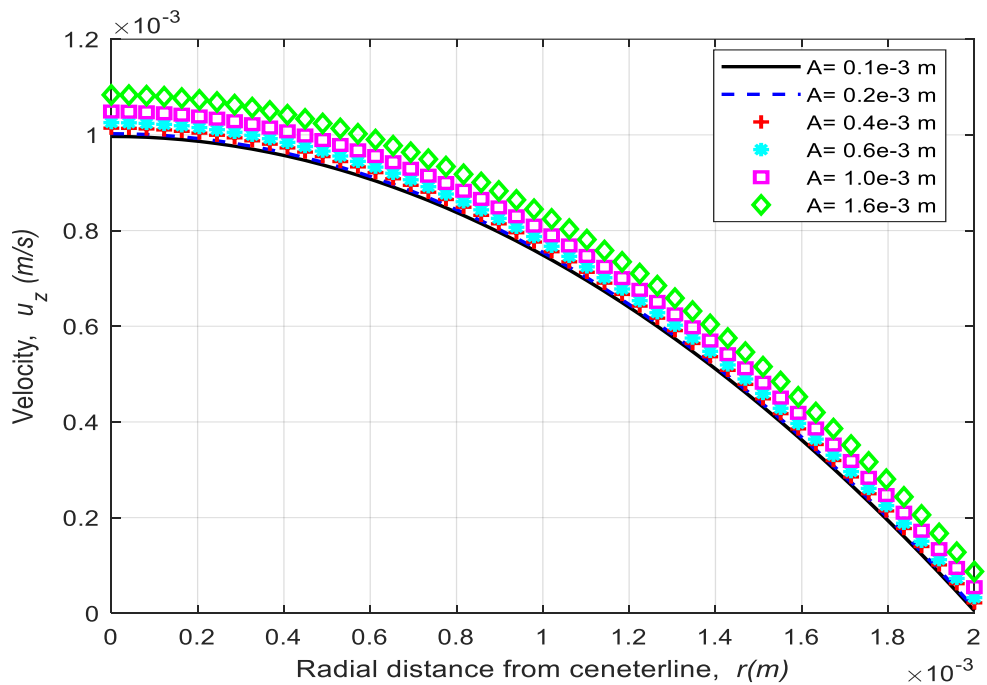

**Fig. 7** Effect of vibration amplitude on the axial velocity profile

Albeit it is important to specify that the curves are less spaced when compared to the plot generated in **Fig. 5**. Inspection of **Fig. 7** suggests that higher vibration amplitude leads to larger axial velocity values along the radial cross-section. The average velocity and instantaneous flowrate as a function of vibration amplitude displayed in **Fig. 8** were then estimated to validate this hypothesis.

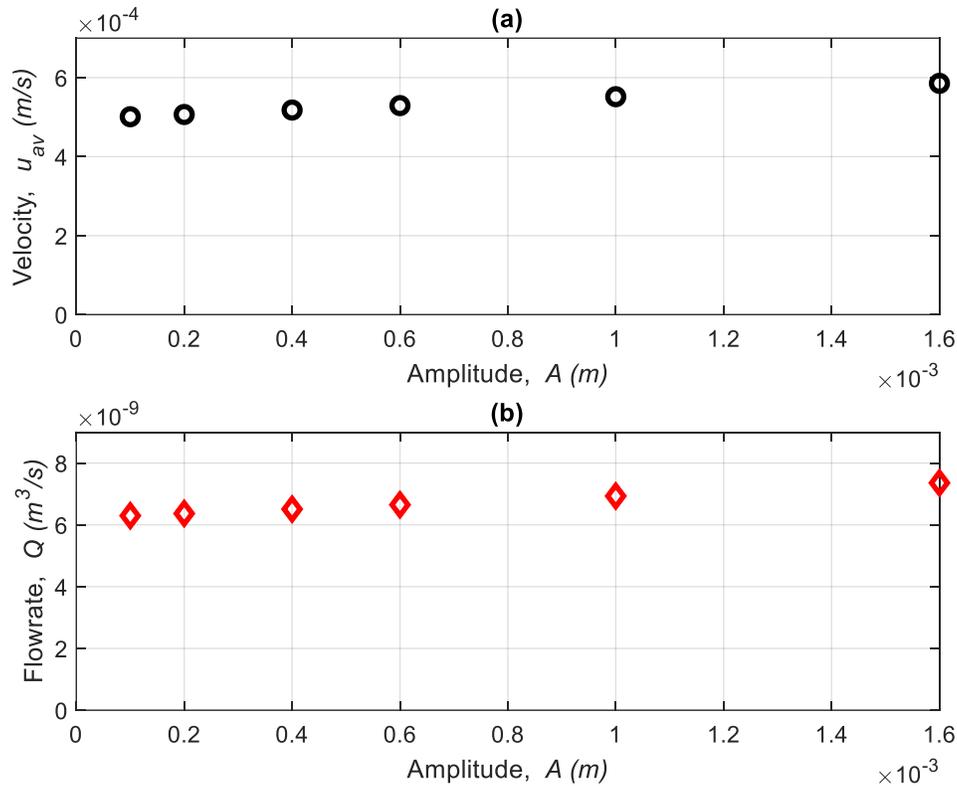

**Fig. 8** Effect of vibration amplitude on (a)average velocity and (b) instantaneous flowrate

**Fig. 8** establishes that both the average velocity and instantaneous flow rate exhibit a positive relationship with vibration amplitude. Specifically, both flow quantities increase with the magnitude of vibration amplitude. Consider **Fig. 8(b)**, the instantaneous flow rate is $7.36 \times 10^{-9}$ m³/s, $8.58 \times 10^{-9}$ m³/s, $9.92 \times 10^{-9}$ m³/s, $11.46 \times 10^{-9}$ m³/s, $13.39 \times 10^{-9}$ m³/s, and $16.10 \times 10^{-9}$ m³/s, for $A = 0.1 \times 10^{-3}$m, $0.2 \times 10^{-3}$m, $0.4 \times 10^{-3}$m, $0.6 \times 10^{-3}$m $1.0 \times 10^{-3}$m, and $1.6 \times 10^{-3}$m, respectively. Finally, the mucus mobilized as a function of vibration amplitude is enumerated in **Table 7**.

**Table 7** Mucus mobilization for different vibration amplitude

| Amplitude (m) | Mean flowrate, $Q_m$ (m$^3$/s) |
|---|---|
| $0.1 \times 10^{-3}$ | $7.27 \times 10^{-9}$ |
| $0.2 \times 10^{-3}$ | $8.10 \times 10^{-9}$ |
| $0.4 \times 10^{-3}$ | $9.77 \times 10^{-9}$ |
| $0.6 \times 10^{-3}$ | $11.43 \times 10^{-9}$ |
| $1.0 \times 10^{-3}$ | $14.77 \times 10^{-9}$ |
| $1.6 \times 10^{-3}$ | $19.77 \times 10^{-9}$ |

**Table 7** confirms that at $f = 15$ Hz, increasing the vibration amplitude by a factor of 2 (i.e., $0.1 \times 10^{-3}$m to $0.2 \times 10^{-3}$m) and 6 (i.e., $0.1 \times 10^{-3}$m to $0.6 \times 10^{-3}$m) leads to 11% and 57% respectively, improvement in mucus mobilization. Therefore, we argue that a significant increase in the amplitude of vibration is required for significant improvement in mucus mobilization.

### 4.2.3. Sensitivity on the mean relaxation time

For viscoelastic fluids, the relaxation time characterizes the time taken for the fluid to return to an unperturbed (i.e., equilibrium) state following a deformation. Typically, large values of relaxation time indicate elastic response while small values imply a viscous response. The role of $\bar{\lambda}_1$ on the flow behavior and mobilization of mucus due to vibration was established by independently varying the magnitude of $\bar{\lambda}_1$ between 0.089 s to 2.67 s (i.e., $\bar{\lambda}_1 < 3$ s) over 4 periods of the cosine wave. The plot depicted in **Fig. 9** indicates that when boundary vibrations are introduced at the tube wall, the curves representing the axial velocity are shifted upwards as $\bar{\lambda}_1$ increases.

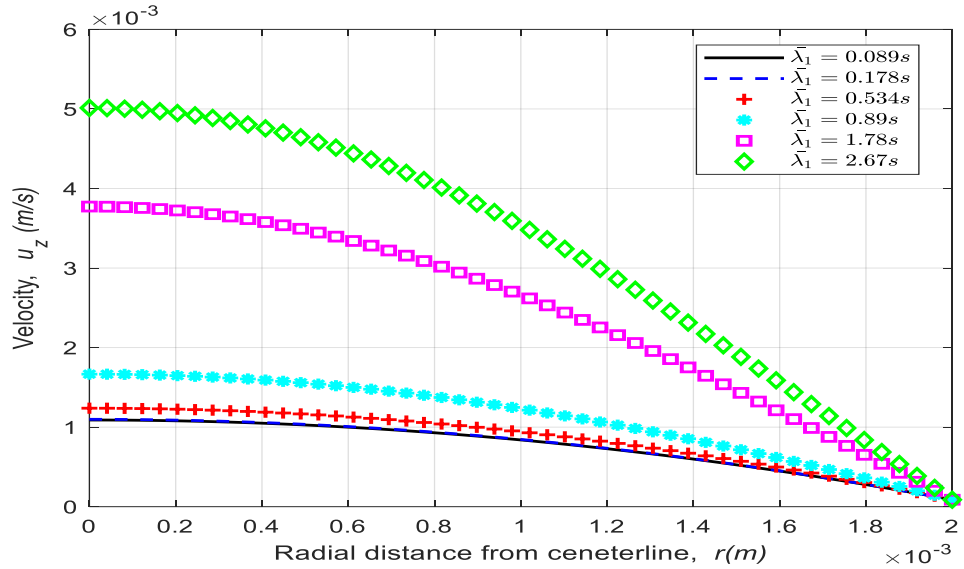

**Fig. 9** Effect of mean relaxation time on the axial velocity

Oscillation at the boundary wall leads to larger values of axial velocity in the tube as the elastic contribution to the overall material response increases. For example, the axial velocity at the pipe centerline equals $1.11 \times 10^{-3}$ m/s, $1.70 \times 10^{-3}$ m/s, and $5 \times 10^{-3}$ m/s for $\bar{\lambda}_1 = 0.089$ s, $0.89$ s, and $2.67$ s, respectively. Therefore, neglecting the elastic contribution from a viscoelastic fluid will lead to underpredicting the axial flow velocity. The relationship between mucus average velocity and instantaneous flow rate as a function of $\bar{\lambda}_1$ are exhibited in **Fig. 10**.

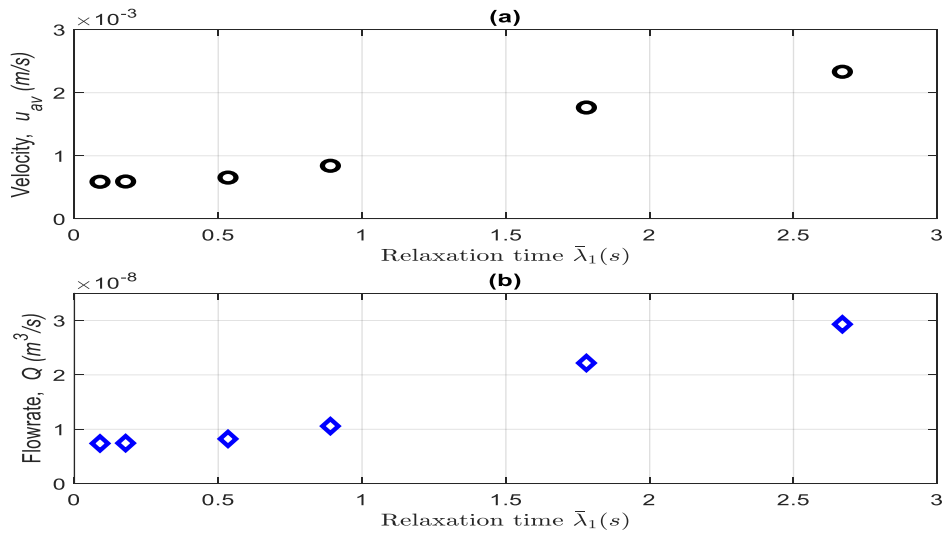

**Fig. 10** Effect of mean relaxation time on (a) average velocity and (b) instantaneous flowrate

Inspection of **Fig. 10** points that the average velocity and instantaneous flowrate exhibit a positive relationship with $\bar{\lambda}_1$. It is evident from **Fig. 10** that the values of average velocity and instantaneous flowrate grow larger as $\bar{\lambda}_1$ increase in magnitude. For example, the average velocity is $0.59 \times 10^{-3}$ m/s, $0.84 \times 10^{-3}$ m/s, and $2.3 \times 10^{-3}$ m/s for $\bar{\lambda}_1 = 0.089$ s, 0.89 s, and 2.67 s, respectively. Similarly, the instantaneous flowrate is $0.74 \times 10^{-8}$ m³/s, $1.06 \times 10^{-8}$ m³/s, and $2.93 \times 10^{-8}$ m³/s, for $\bar{\lambda}_1 = 0.089$ s, 0.89 s, and 2.67 s, respectively. **Table 8** summarizes the mucus mobilized in the vibrated tube for the given values of $\bar{\lambda}_1$.

**Table 8** Mucus mobilization at the different mean relaxation time

| Relaxation time, $\bar{\lambda}_1$(s) | Mean flow rate, $Q_m$ (m³/s) |
|---|---|
| 0.089 | $2.17 \times 10^{-8}$ |
| 0.178 | $2.37 \times 10^{-8}$ |
| 0.534 | $3.21 \times 10^{-8}$ |
| 0.89 | $4.05 \times 10^{-8}$ |
| 1.78 | $6.13 \times 10^{-8}$ |
| 2.76 | $8.12 \times 10^{-8}$ |

**Table 8** indicates that for a given vibration frequency and amplitude (i.e., 15 Hz and $1.6 \times 10^{-3}$ m), a larger value of mucus is mobilized within the tube as the magnitude of $\bar{\lambda}_1$ increases. Specifically, a 9% and 48% improvement in the mean flow rate was observed when $\bar{\lambda}_1$ was increased from 0.089 s to 0.178 s and from 0.089 s to 0.534 s, respectively. Therefore, we argue that as the elastic contribution of mucus increases due to chronic lung infection or inflammation, the mucus clearance efficiency can be improved via introducing boundary oscillations.

### 4.2.4. *Sensitivity on the zero-shear viscosity*

Viscosity is a measure of the resistance to deformation of fluid at the flow condition and it quantifies material response for viscous fluids. It is important to recall that for the linear viscoelastic theory, the viscosity is independent of shear rate and is quantified by the zero-shear viscosity. The axial velocity profile as a function of radial distance after 4 periods of the cosine wave is illustrated in **Fig. 11**.

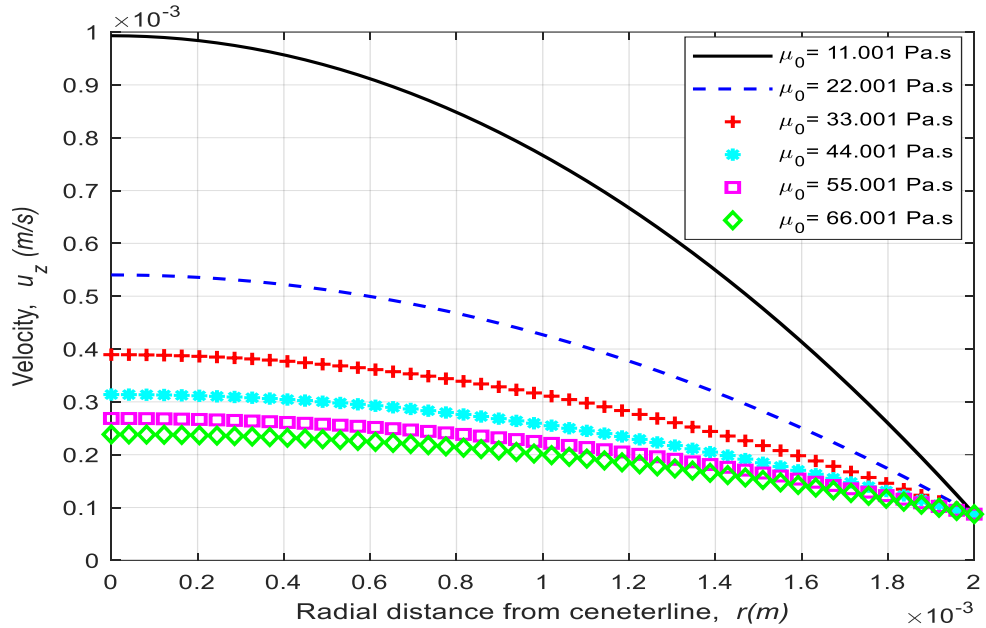

**Fig. 11** Effect of zero-shear viscosity on the axial velocity profile

As shown in **Fig. 11**, the curves are shifted downwards as the magnitude of $\mu_0$ increases. Take, for example, the axial velocity at the tube centerline equals $0.931 \times 10^{-3}$ m/s, $0.314 \times 10^{-3}$ m/s, and $0.238 \times 10^{-4}$ m/s, for $\mu_0 = 11.001$ Pa·s, $44.001$ Pa·s, and $66.001$ Pa·s, respectively. Therefore, the mucus transport within the tube decreases as the mucus viscosity increases. To further validate this hypothesis, the average velocity and instantaneous flowrate of mucus within the tube as a function of $\mu_0$ are shown in **Fig. 12**.

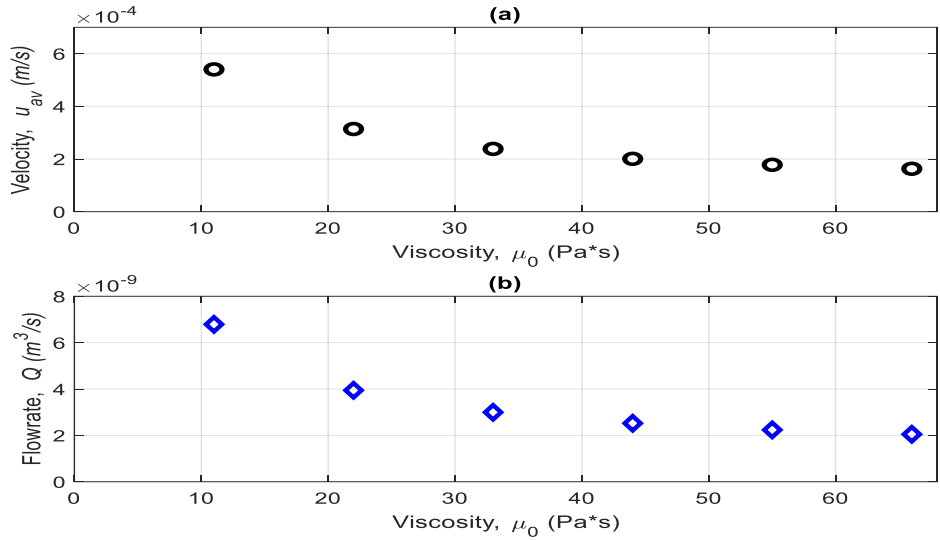

**Fig. 12** Effect of zero-shear viscosity on (a) average velocity and (b) volumetric flowrate

Observation of **Fig. 13** confirms that both the average velocity and instantaneous flow rate display a negative trend with $\mu_0$. Specifically, we note that the average velocity and instantaneous flowrate diminish as viscosity magnitude increases. For instance, the average velocity is $5.40 \times 10^{-4}$ m/s, $2.01 \times 10^{-4}$ m/s, and $1.63 \times 10^{-4}$ m/s for $\mu_0 = 11.001$ Pa.s, 44.001 Pa.s, and 66.001 Pa.s, respectively. While the instantaneous flowrate is $6.79 \times 10^{-9}$ m³/s, $2.52 \times 10^{-9}$ m³/s, and $2.05 \times 10^{-9}$ m³/s, for $\mu_0 =$ for 11.001 Pa.s, 44.001 Pa.s, and 66.001 Pa.s, respectively. The effect of mucus viscosity on mucus mobilization (i.e., mean flow rate) after four periods of the cosine wave is quantified in **Table 8**.

**Table 9** Mucus mobilization for different zero-shear viscosity values

| Zero-shear viscosity, $\mu_0$ (Pa.s) | Mean flow rate, $Q_m$ (m³/s) |
|---|---|
| 11.001 | $1.92 \times 10^{-8}$ |
| 22.001 | $1.63 \times 10^{-8}$ |
| 33.001 | $1.53 \times 10^{-8}$ |
| 44.001 | $1.48 \times 10^{-8}$ |
| 55.001 | $1.45 \times 10^{-8}$ |
| 66.001 | $1.43 \times 10^{-8}$ |

**Table 9** indicates that for a given value of vibration frequency of 15 Hz and amplitude of $1.6 \times 10^{-3}$ m, a 15% and 26% decrease in mucus mobilization were estimated by increasing the viscosity by a factor of 2 (i.e. 11 Pa.s to 22 Pa.s) and 6 (i.e., 11 Pa.s to 66 Pa.s), respectively.

## 5. CONCLUSION

In this paper, a semi-analytical solution for the Poiseuille flow of mucus in a vibrated cylindrical tube was derived by adopting the Laplace transform method on the governing partial differential equations. The derived expressions for the velocity field, average velocity, instantaneous flowrate, and mean flowrate were subsequently utilized to examine the transient flow response and mobilization of mucus due to the sinusoidal oscillations at the tube wall. It is important to highlight some of the key advantages of the proposed analytical solution includes its convenience for rapid prototyping due to speed, stability, and accuracy since numerical solutions for viscoelastic flows sometimes fail to converge especially for high Weissenberg numbers. Besides, the presented analytical solution provides a convenient tool for experimenting with different wall velocity expressions (i.e., sinusoidal, non-sinusoidal, and other novel functions). The only condition is that the proposed wall velocity function satisfies the conditions for applicability of Laplace transform (i.e., the function must be locally integrable for the interval $[0, \infty]$). Furthermore, the analytical solution allows for studying viscoelastic flows driven by time-dependent pressure gradients which may stem from airflow due to coughing [82], and pulsating pressure gradients. The main findings from this research are outlined as follows:

1. The vibration frequency exhibits a positive relationship with both the average velocity and instantaneous flow rate. Analysis indicated that the mucus mobilized in the tube improved by 68% and 343% when the vibration frequency was increased by a factor of 2 (i.e., 15 Hz to 30 Hz) and 6 (i.e., 15 Hz to 90 Hz), respectively.

2. The vibration amplitude displays a positive trend with the average velocity and instantaneous flowrate. Besides, mucus mobilization increased by 11% and 57% when the vibration amplitude by a factor of 2 (i.e., $0.1 \times 10^{-3}$m to $0.2 \times 10^{-3}$m) and 6 (i.e., $0.1 \times 10^{-3}$m to $0.6 \times 10^{-3}$m), respectively.

3. The mean relaxation time depicts a positive correlation with the average velocity and instantaneous flow rate. Results indicated that for a frequency of 15 Hz and vibration amplitude of $1.6 \times 10^{-3}$ m, mucus mobilization improved by 9% and 48% when the value of $\bar{\lambda}_1$ was increased by a factor of 2 (i.e., 0.089 s to 0.178 s) and a factor of 6 (i.e., 0.089 s to 0.534 s), respectively.

4. The average velocity and instantaneous flow display a negative trend with the zero-shear viscosity. Numerical experiments indicated that mucus mobilization decreased by 15% and 26% when the zero-shear viscosity was multiplied by a factor of 2 (i.e. 11 Pa.s to 22 Pa.s) and 6 (i.e., 11 Pa.s to 44 Pa.s), respectively.

## 6. ACKNOWLEDGEMENTS

The authors would like to acknowledge the support provided by Dymedso Inc, Natural Sciences and Engineering Research Council of Canada and by a grant in aid of research from the Mitacs Accelerate Fellowship.

## 7. NOMENCLATURE

| | |
|---|---|
| $A$ | The amplitude of wall displacement, m |
| $a_1$ | coefficient defined by terms in Eq. (A-16) associated with boundary conditions |
| $a_2$ | coefficient defined by terms in Eq. (A-16) associated with boundary conditions |
| $b_0$ | Acceleration coefficient, m/s$^2$ |
| $b_1$ | Phase angle, dimensionless |
| $c_3$ | coefficient, Eq. (A-25) |
| $D$ | Diameter of pipe, m |
| $e_{ij}$ | Defined in Eq. (25) |
| $f(t)$ | General time function for wall velocity |
| $F(s)$ | Laplace transform of $f(t)$ |
| $g$ | Gravity constant, 9.81 m/s$^2$ |
| $G$ | Elastic modulus, Pa |
| $h(s)$ | The Oldroyd-B shear stress transfer function |
| $I_0$ | Modified Bessel function of the first kind of order 0 |
| $I_1$ | Modified Bessel function of the first kind of order 0 |
| $j$ | notation for $\sqrt{-1}$ |
| $J_0$ | Bessel function of order 0 |
| $J_1$ | Bessel function of order 1 |
| K | Defined in Eq. (21) |
| $K_0$ | Modified Bessel function of the second kind of order 0 |
| $K_n$ | Stehfest Weighting coefficient, Eq. (22) |
| $L$ | Length of pipe, m |
| $m$ | Mode/branch of Oldroyd-B model |
| $\dot{m}$ | Mass flow rate, kg/s |
| $p$ | Pressure of the pipe, Pa |
| $P_A$ | Pressure at the upstream end of the pipe, Pa |
| $P_B$ | Pressure at the downstream end of the line, Pb |
| $r$ | Radial position from the center of the fluid pipe, m |

$Re_v = \dfrac{A\omega D}{v_e}$      Vibrational Reynolds number, dimensionless

$Q$      Instantaneous flow rate, m$^3$/s

$\hat{Q}$      Laplace transform of instantaneous flow rate

$Q_{HP}$      Instantaneous flow rate at stationary condition, m$^3$/s

$Q_m$      Mean flowrate defined in Eq. (20), m$^3$/s

$\hat{Q}_m$      Laplace transform of mean flowrate

$R$      the radius of the pipe, m

$s$      Laplace transform parameter

$S$      Defined in Eq. (25)

$t$      Elapsed time, s

$T$      Elapsed time at end of the period, s

$u_{av}$      Average velocity, m/s

$u_W$      Wall velocity, m/s

$u_z$      fluid velocity a function of $r$, m/s

$\hat{U}_z$      Laplace transform of the fluid velocity

$\hat{U}_W$      Laplace transform of wall velocity

$z$      axial position along the fluid line, m

**Greek Symbols**

$\alpha$      Giesekus parameter, dimensionless

$\rho$      fluid density, kg/m$^3$

$\mathcal{L}$      Laplace transform operator defined in Eq. (A-1)

$\lambda_1$      Relaxation time constant, s

$\bar{\lambda}_1$      Mean relaxation time constant, s

$\lambda_2$      Retardation time constant, s

$\bar{\lambda}_2$      Mean retardation time constant, s

$\sigma$      Shear stress, N/m$^2$

$\overset{\triangledown}{\sigma}_m$      The upper convected time derivative of the stress tensor

$\mu_0$      Zero shear rate viscosity, Pa·s

$\mu_s$      Solvent component of viscosity, Pa·s

$\mu_p$      The polymer component of viscosity, Pa·s

$\mu$      Dynamic viscosity, Pa·s

$v_e$      Kinematic viscosity, m$^2$/s

$\omega$      Angular frequency, rad/s

$\psi = -\dfrac{\partial p}{\partial z} + \rho g$      Net pressure gradient

# 8. APPENDIX A: DERIVATION OF ANALYTICAL SOLUTION

The analytical solution to the flow in a vibrated tube is arrived at employing the Laplace transform method defined by the expression

$$\mathcal{L}[f(t)] = F(s) = \int_0^\infty f(t)e^{-st}dt \tag{A-1}$$

Where $\mathcal{L}$ and $s$ are the Laplace transform operator and Laplace variable.

The Laplace transform of Eqs. (3) and (4) respectively yields

$$\rho\left[s\hat{U}_z - u_z(0)\right] = \hat{\psi} + \frac{1}{r}\frac{\partial}{\partial r}(r\hat{\sigma}) \tag{A-2}$$

$$\frac{\partial \hat{U}_z}{\partial z} = 0 \tag{A-3}$$

where $\hat{U}_z, \hat{\psi} = \frac{\psi}{s}$ and $\hat{\sigma}$ are the axial velocity field, pressure gradient, and stress tensor in the Laplace domain respectively.

Given the assumption of zero initial condition i.e. $u_z(0) = 0$, Eqs. (A-2) and (A-3) reduces to

$$\rho s\hat{U}_z = \hat{\psi} + \frac{1}{r}\frac{\partial}{\partial r}(r\hat{\sigma}) \tag{A-4}$$

$$\frac{\partial \hat{U}_z}{\partial z} = 0 \tag{A-5}$$

Applying the operator $\mathcal{L}$ on Eq. (8) and simplifying leads to

$$\hat{\sigma} = \mu_0 h(s)\frac{d\hat{U}_z}{dr} \tag{A-6}$$

where $h(s)$ denotes the normalized Oldroyd-B shear stress transfer function defined by

$$h(s) = \left[\frac{1+\lambda_2 s}{1+\lambda_1 s}\right] \tag{A-7}$$

It is important to note the following:

1) At $\lambda_2 = 0$ the stress-shear rate constitutive relation reduces to the Maxwell model.
2) At $\lambda_2 = \lambda_1$ the stress shear rate constitutive equation reduces to a Newtonian fluid with viscosity $\mu_0$.

To arrive at the expression for the flow field, we combine the momentum equation with the Oldroyd-B stress relation as follows

$$\rho s\hat{U}_z = \hat{\psi} + \frac{1}{r}\frac{\partial}{\partial r}\left[r\mu_0 h(s)\frac{\partial \hat{U}_z}{\partial r}\right] \tag{A-8}$$

Adopting the chain rule of differentiation, Eq. (A-8) yields

$$\rho s \hat{U}_z - \hat{\psi} = \mu_0 h(s) \left[ \frac{\partial^2 \hat{U}_z}{\partial r^2} + \frac{1}{r} \frac{\partial \hat{U}_z}{\partial r} \right] \tag{A-9}$$

Re-arranging Eq. (A-9) and aggregating similar terms yields

$$\frac{\partial^2 \hat{U}_z}{\partial r^2} + \frac{1}{r} \frac{\partial \hat{U}_z}{\partial r} = \frac{1}{\mu_0 h(s)} \left[ \rho s \hat{U}_z - \hat{\psi} \right] \tag{A-10}$$

Simplifying Eq. (A-10) leads to the expression

$$\frac{\partial^2 \hat{U}_z}{\partial r^2} + \frac{1}{r} \frac{\partial \hat{U}_z}{\partial r} - \frac{\rho s}{\mu_0 h(s)} \hat{U}_z = -\frac{\hat{\psi}}{\mu_0 h(s)} \tag{A-11}$$

Eq. (A-11) can be re-written as follows

$$\frac{\partial^2 \hat{U}_z}{\partial r^2} + \frac{1}{r} \frac{\partial \hat{U}_z}{\partial r} - c^2 \hat{U}_z = -\frac{\hat{\psi}}{\mu_0 h(s)} \tag{A-12}$$

where

$$c^2 = \frac{\rho s}{\mu_0 h(s)} = \frac{s}{v_e h(s)} \tag{A-13}$$

Inspection of Eq. (A-12) reveals a second-order non-homogenous differential equation with a general solution comprising the complementary and particular solutions as follows:

$$\hat{U}_z = \hat{U}_{zC} + \hat{U}_{zP} \tag{A-14}$$

The complementary solution is obtained from the homogenous equation

$$\frac{\partial^2 \hat{U}_z}{\partial r^2} + \frac{1}{r} \frac{\partial \hat{U}_z}{\partial r} - c^2 \hat{U}_z = 0 \tag{A-15}$$

The Eq. (A-16) is a modified Bessel equation with a solution

$$\hat{U}_{zc} = a_1 \text{I}_0(cr) + a_2 \text{K}_0(cr) \tag{A-16}$$

Where $a_1$ and $a_2$ are coefficients to be obtained from the specified boundary conditions, and $j$ is the notation for $\sqrt{-1}$, $\text{I}_0(.)$ and $\text{K}_0(.)$ are the modified Bessel functions of the first and second kind with order 0.

Employing the method of undetermined coefficients, the particular solution of Eq. (A-14) yields

$$\hat{U}_{zP} = \frac{\hat{\psi}}{\rho s} \tag{A-17}$$

Substituting Eqs. (A-16) and (A-17) into Eq. (A-14) leads to

$$\hat{U}_z = a_1 \text{I}_0(cr) + a_2 \text{K}_0(cr) + \frac{\hat{\psi}}{\rho s} \tag{A-18}$$

The boundary conditions in Laplace space can be expressed as follows:

$$\hat{U}_z = \hat{U}_W \quad \text{at } r = R \tag{A-19}$$

And

$$\frac{\partial \hat{U}_z}{\partial r} = 0 \quad \text{at } r = 0 \text{ for all } t \tag{A-20}$$

Recall the following properties of the modified Bessel functions

$$\frac{\partial \text{K}_0(cr)}{\partial r} = -c \text{K}_1(cr) \tag{A-21}$$

Additionally, we have that

$$\frac{\partial I_0(cr)}{\partial r} = cI_1(cr) \tag{A-22}$$

Therefore, applying the operator $\frac{d}{dr}$ on Eq. (A-18) results in

$$\frac{\partial \hat{U}_z}{\partial r} = a_1 cI_1(cr) - a_2 cK_1(cr) \tag{A-23}$$

Employing the boundary condition at $r = 0$ in Eq. (A-23) yields

$$a_2 = 0 \tag{A-24}$$

Therefore, Eq. (A-18) simplifies to:

$$\hat{U}_z = a_1 I_0(cr) + \frac{\hat{\psi}}{\rho s} \tag{A-25}$$

Similarly, at $r = R$, combining Eq. (A-25) and Eq. (A-19) leads to the expression

$$\hat{U}_W = a_1 I_0(cR) + \frac{\hat{\psi}}{\rho s} \tag{A-26}$$

Therefore, the coefficient, $a_1$ is defined as

$$a_1 = \frac{1}{I_0(cR)}\left(\hat{U}_W - \frac{\hat{\psi}}{\rho s}\right) \tag{A-27}$$

## 10. Figure Caption

1. **Fig. 1** Schematic of the flow geometry
2. **Fig. 2** Meshed 2D axisymmetric geometry of the vibrated tube

## 11. Table Caption